# Anomalous Nernst Effect in Ferromagnetic Weyl Semimetal

*Udai Prakash Tyagi and *Partha Goswami*
*D.B.College, University of Delhi, Kalkaji, New Delhi-110019, India*
*Email of the first author: uptyagi@yahoo.co.in*
*\*Email of the corresponding author: physicsgoswami@gmail.com*

**Abstract**

In a three-dimensional Dirac semimetal the time reversal symmetry (TRS) or the inversion symmetry (IS) is not broken. With either of these symmetries broken, the Dirac points in the three-dimensional band structure split up into pairs of so-called Weyl points. The ferromagnetic Weyl semimetals (FMWSM), such as $Co_3Sn_2S_2$, feature pairs of Weyl points characterized by the opposite chiralities. In this communication we study FMWSM based on TRS broken continuum and lattice Hamiltonians. The latter one is more realistic and represents $Co_3Sn_2S_2$. These models include all essential ingredients leading to the formation of a pair of Weyl nodes and tilted Weyl cones. Our analysis shows a large anomalous Nernst conductivity which is unlocked due to the divergent Berry curvature - a local manifestation of the geometric properties of electronic wavefunctions - at Weyl points.

**1. Introduction**

The Weyl fermion quasiparticle( WFQPs)are massless and have a definite chirality- a new degree of freedom in the relativistic Hamiltonian describing a relativistic quantum mechanical state. For such QPs, the chirality/handedness is determined by the sign of its spin polarization along the momentum direction **[1,2]**. One may choose the index $\chi = \pm 1$ to represent two Weyl fermions (WFs) of opposite chiralities, say at wavevectors $+k_0$ and $-k_0$, in a Weyl semimetal (WSM). This host material is such that the time reversal symmetry(TRS) is broken but the parity is preserved. A gauge field that couples differently to WFs in this material with two chiralities is called the chiral gauge field. As a consequence, the chiral charge conservation is violated in a theory of WFQPs, typically in the presence of parallel electric and magnetic fields. The violation is referred to as the chiral anomaly**[3].**The WF and the chiral anomaly **[4,5]** act as a link between the condensed matter and high-energy physics (HEP). The introduction of the idea was originally motivated by the decay process of neutral pion in the context of HEP **[5]**. These quasi-particles play a central role in the relativistic field theory. In condensed matter physics (CMP), the central concepts of causality and the allied event horizon in spacetime can be carried over into the field of correlated WSM. This strong association**[6]** between condensed matter and high-energy physics is depicted in Figure 1(a).

The Weyl semimetals,e.g., transition mono-pnictides TaAs,TaP, NbP, etc.**[7– 12]**, WTe$_2$ **[13]**, magnetic compounds such as $Co_3Sn_2S_2$ **[14 -21]**, $Co_2MnGa$ **[22]**, and so on have been one of the areas of great interest of the CMP community for quite some time. These materials provide the realization of WFQPs -- low energy excitations - around the Weyl points. The Weyl nodes(WNs) are the isolated points in the Brillouin zone (BZ) carrying non-trivial chiral charge ( chiral charge is equal to the Chern number on a small spherical manifold enclosing the Weyl point in the bulk BZ) where pairs of nondegenerate bands of opposite chirality cross/ touch each other linearly or almost linearly. These points are seperated in the momentum space. While a right-handed WN ( chirality index $\chi = +1$) is a source, a left-handed one ( chirality index $\chi = -1$) is a drain (see Figure 1(b)) in the context of Berry Curvature(BC). It may be mentioned that the presence of WNs is manifested by non-trivial topological properties such as negative magnetoresistance ascribed to chiral anomaly and the exotic surface Fermi arcs (SFA). As shown in Figure 1(b), SFA is an open curve joining the projections of WNs of opposite chiralities in the bulk BZ onto the surface BZ. The chiral charge mentioned above is responsible for the existence of the SFA**[23,24]**. More about SFA is discussed in section 5. The non-trivial topology of the WSMs**[7-21,25]** is an upshot of the broken time reversal symmetry (TRS) or the space inversion symmetry (SIS) or the both of them with/without strong correlations.

The stability of the Weyl nodes requires the broken symmetry scenario. If TRS and SIS are both broken the minimal number of Weyl nodes must be two. This is in agreement with the Nielsen-Ninomiya theorem**[26]** that requires the number of Weyl points in the Brillouin zone to be even. Quite intriguingly, it has been shown that the ferromagnetic WSM $Fe_3Sn_2$ **[27-29]** shows a large, albeit even number of Weyl nodes, close to the Fermi energy. Such host systems have been reported to provide favorable prospect for the emergence of interesting quantum phases/ phase transition, such as the Weyl charge density wave (CDW)(standing waves with a periodicity different from the underlying atomic lattice) phase in $(TaSe_4)_2I$ and $Mo_3Al_2C$ **[6,30],** the ferromagnetic ground state and the possible nontrivial superconductor state of the alluaudite type compound $K_2Mn_3(AsO_4)_3$**[31]**, the interesting transition from the normal CDW phase to exotic, strong coupling superconducting state possibly with s-wave symmetry in kagome metals$AV_3Sb_5$(A=K,Rb,Cs)**[32-34]**, the ferromagnetic Weyl semi-metallic (FMWSM) phase of the transition metal based Kagome compound $Co_3Sn_2S_2$ **[14 -21],** and so on. In fact, in Kagome materials the atoms in a layer form a lattice resembling a traditional Japanese basket-weaving pattern.The study of these materials,including $K_2Mn_3(AsO_4)_3$**[31]** and the frustated Kagome ferromagnetic alloy $Fe_3Sn_2$ **[27-29]**, provides an attainable pulpit to probe the intrinsic properties related to Weyl nodes, and the allied device applications. In this paper, however, we theoretically investigate continuum and lattice models of FMWSM, such as the Kagome compound $Co_3Sn_2S_2$**[14-21]**. The characteristic feature of FMWSM is a large anomalous Nernst effect(ANE) **[35]** which is unlocked due to the divergent BC - a local manifestation of the geometric properties of electronic wavefunctions - at Weyl nodes. We obtain the Berry curvature and the anomalous Nernst conductivity within the framework of a continuum model **[16, 20,21]** of FMWSM in this communication.

The paper is organized as follows: In section 2, we present a continuum model and the more realistic lattice model **[16,21]** of FMWSM bulk. The former includes all essential ingredients leading to the formation of a pair of Weyl nodes and tilted Weyl cones. We obtain the energy eigenspectra and the corresponding eigenvectors of these model Hamiltonians in section 3. In section 4, we present our findings on the topological properties appearing in the continuum model in the section 2. Finally, in Section 5, we present discussion and conclusion.
.

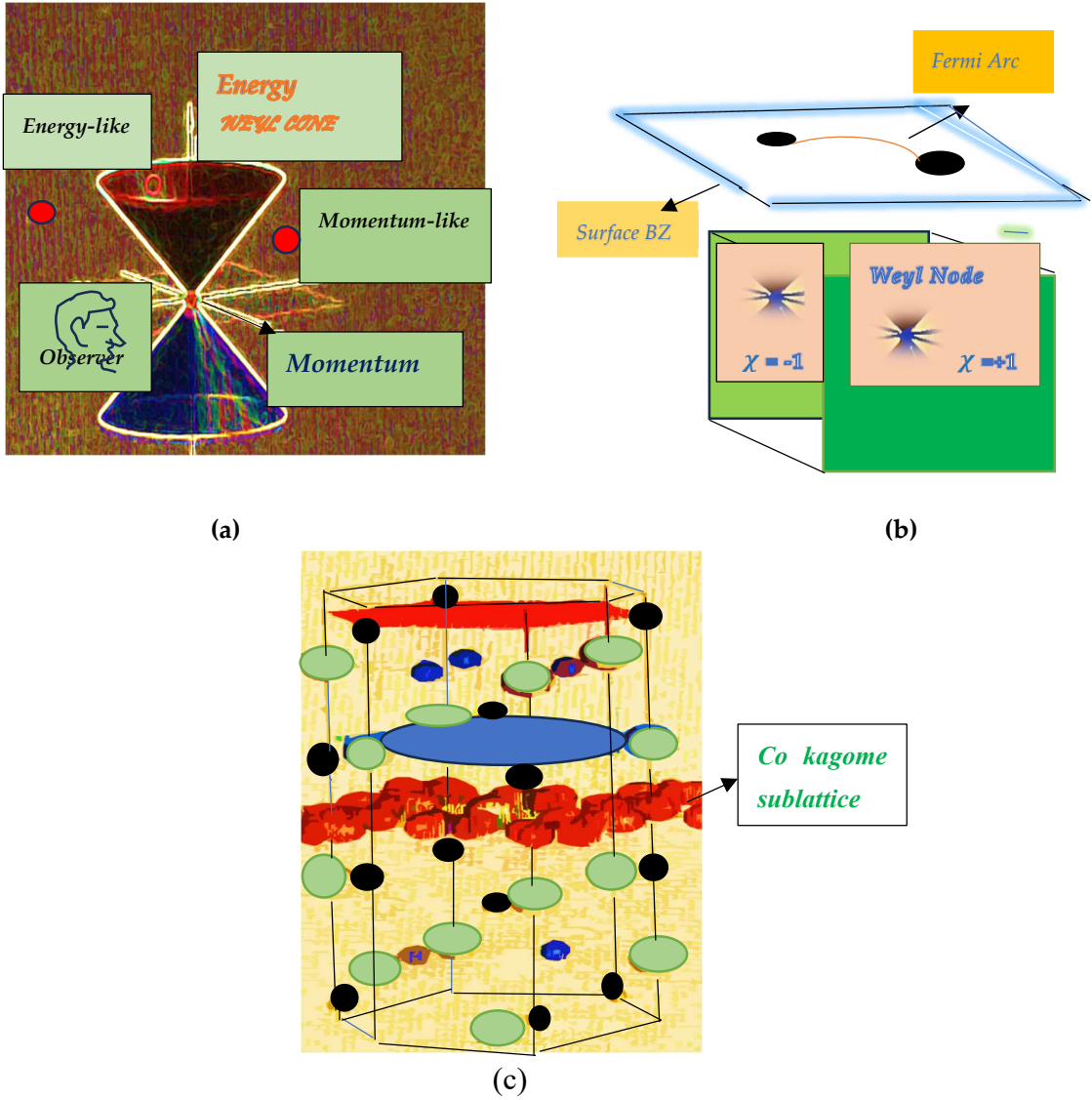

**Figure 1. (a)** This caricature cartoon represents how the central concepts of causation and the consorted event horizon in spacetime can be carried over into the field of correlated Weyl materials, and thus depicting strong association between condensed matter and high-energy physics. **(b)** A cartoon of the Weyl nodes (WNs) of different chirality in the bulk Brillouin zone(BZ) and the Fermi arc (FA)in the surface BZ (for the WSM case). The presence of WN is manifested by surface FA that connect the WNs of opposite chirality. The chirality index $\chi = +1(-1)$ corresponds to momentum space pseudo-magnetic monopole(anti-monopole). The former is a source of Berry curvature (BC), while the latter is a drain. (c) The shandite– type crystal structure of $Co_3Sn_2S_2$ with Sn (green), S (black), and Co (blue) atoms. The colored plane marks are blue for the Sn–type, red for the S–type, and Sn–type (blue).

## 2. Model Hamiltonians

### A. High Symmetry Points

The Weyl fermion quasiparticles in a Weyl semimetal (WSM) are very robust. Except the translation symmetry of the crystal lattice, these fermions do not depend on other symmetries . These fermions possess a high degree of mobility. As regards FMWSM, the representative compound $Co_3Sn_2S_2$ is a ferromagnet with a Curie temperature of $T_c \sim$ 180K. The structure is composed of triangular Sn and S layers interspersed between the Kagome lattice planes of Co atoms(see Figure 1(c)). These Co atoms are responsible for ferromagnetic order. Ozawa et al. have analysed a momentum-space model Hamiltonian **[16,21,35]** of $Co_3Sn_2S_2$ and the corresponding electronic band structure. This two-orbital archetypal model **[16,21]** represents FMWSM $Co_3Sn_2S_2$ with great details with the effective Kane-Mele type spin-orbit coupling (SOC)**[36]** between the Co atoms in the Kagome plane. The unit cell includes three Co atoms on the Kagome lattice and one Sn atom on the triangular lattice. The model is described by a primitive lattice with three basis vectors: $a_1$, $a_2$, and $a_3$ where $a_1 = a\left(\frac{1}{2}, 0, \frac{c}{a}\right)$, $a_2 = a\left(-\frac{1}{4}, \frac{\sqrt{3}}{4}, \frac{c}{a}\right)$ and $a_3 = a\left(-\frac{1}{4}, -\frac{\sqrt{3}}{4}, \frac{c}{a}\right)$. In the case of bulk $Co_3Sn_2S_2$ the lattice parameters are 5.37 Å and 13.17 Å along ($a$, $b$) and $c$ directions, respectively. The lattice parameters a, b, and c are measured by single crystal diffraction to be 5.3657 Å, 5.3657 Å and 13.1710 Å, respectively. While he angle α = 59.91647° (58.33°), the angle is $\gamma = 118.78°$. The intralayer lattice vectors of a Kagome layer ($b_1$, $b_2$, $b_3$) (the first-nearest-neighbour vectors),($d_1$, $d_2$, $d_3$) (the second-nearest-neighbour vectors), and the interlayer lattice vectors of Kagome layers ($c_1$, $c_2$, $c_3$) (the first-nearest-neighbour vectors) were represented in ref.**[16,21,35]** in a certain way. In this communication, we, however, choose to write

$$a_1 = a\left(\cos\frac{\gamma}{2}, 0, \frac{c}{a}\right), a_2 = a\left(-\frac{1}{2}\cos\frac{\gamma}{2}, \frac{\sqrt{3}}{2}\cos\frac{\gamma}{2}, \frac{c}{a}\right),$$

$$a_3 = a\left(-\frac{1}{2}\cos\frac{\gamma}{2}, -\frac{\sqrt{3}}{2}\cos\frac{\gamma}{2}, \frac{c}{a}\right), \quad (1)$$

where the interaxial angle γ is that between $a_1$ and $a_2$. The lattice vectors ($b_1$, $b_2$, $b_3$),($d_1$, $d_2$, $d_3$), and ($c_1$, $c_2$, $c_3$) are expressed in terms of the angle γ as shown below:

$$b_1 = \frac{a_2 - a_1}{2} = \frac{a}{2}\left(-\frac{3}{2}\cos\frac{\gamma}{2}, \frac{\sqrt{3}}{2}\cos\frac{\gamma}{2}, 0\right), b_2 = \frac{a_3 - a_2}{2} = \frac{a}{2}\left(0, -\sqrt{3}\cos\frac{\gamma}{2}, 0\right), b_3 = \frac{a_1 - a_3}{2}$$

$$\frac{a}{2}\left(\frac{3}{2}\cos\frac{\gamma}{2}, \frac{\sqrt{3}}{2}\cos\frac{\gamma}{2}, 0\right), \quad (2)$$

$$d_1 = b_3 - b_2 = \frac{a}{2}\left(\frac{3}{2}\cos\frac{\gamma}{2}, \frac{3\sqrt{3}}{2}\cos\frac{\gamma}{2}, 0\right), d_2 = b_1 - b_3 = \frac{a}{2}\left(-3\cos\frac{\gamma}{2}, 0, 0\right),$$

$$d_3 = b_2 - b_1 = \frac{a}{2}\left(\frac{3}{2}\cos\frac{\gamma}{2}, \frac{-3\sqrt{3}}{2}\cos\frac{\gamma}{2}, 0\right), \quad (3)$$

$$c_1 = \frac{a_1 + a_2}{2} = \frac{a}{2}\left(\frac{1}{2}\cos\frac{\gamma}{2}, \frac{\sqrt{3}}{2}\cos\frac{\gamma}{2}, 2\frac{c}{a}\right), c_2 = \frac{a_2 + a_3}{2} = \frac{a}{2}\left(-\cos\frac{\gamma}{2}, 0, 2\frac{c}{a}\right),$$

$$c_3 = \frac{a_3 + a_1}{2} = \frac{a}{2}\left(\frac{1}{2}\cos\frac{\gamma}{2}, -\frac{\sqrt{3}}{2}\cos\frac{\gamma}{2}, 2\frac{c}{a}\right). \quad (4)$$

The reciprocal space is very useful as it provides a natural language for describing periodic systems and all periodic responses are contained within the first Brillouin zone (BZ). The reciprocal lattice vectors ($\boldsymbol{\rho}_1(\gamma), \boldsymbol{\rho}_2(\gamma), \boldsymbol{\rho}_3(\gamma)$) in terms of the angle γ are given by $\frac{a\boldsymbol{\rho}_1(\gamma)}{2\pi} = (4\sqrt{3}\zeta, 2\zeta, 3\xi)$, $\frac{a\boldsymbol{\rho}_2(\gamma)}{2\pi} = (-2\sqrt{3}\zeta, 6\zeta, 2\xi)$, and $\frac{a\boldsymbol{\rho}_3(\gamma)}{2\pi} = (-2\sqrt{3}\zeta, -8\zeta, 2\xi)$, where $\zeta \equiv$ $\frac{1}{7\sqrt{3}} \sec\frac{\gamma}{2}$, and $\xi = \frac{a}{7c}$. By using these reciprocal vectors, it is easy to translate the lattice points in reciprocal space and build the corresponding reciprocal lattice for the prototypical kagome magnetic WSM under consideration. The high-symmetry points (HSPs) in the reciprocal space are important for describing the electronic and magnetic properties of solids. Upon using the frame-work outlined in ref.[16,21], the coordinates of high symmetry k-points have been obtainted here. The relevant HSPs, viz. $T, L, U,$ and $W$ points are given by $\left(\frac{2\pi}{a}\right)\left(0,0,\frac{7\xi}{2}\right)$, $\left(\frac{2\pi}{a}\right)\left(2\sqrt{3}\zeta, \zeta, \frac{3\xi}{2}\right)$, $\left(\frac{2\pi}{a}\right)\left(\eta, 0, \frac{7\xi}{2}\right)$, and $\left(\frac{2\pi}{a}\right)\left(-\eta, \frac{\eta}{\sqrt{3}}, \frac{7\xi}{2}\right)$, respectively, where $\eta \equiv 2\sqrt{3}\,\zeta - \frac{\sqrt{3}\xi^2}{2\zeta}$. We calculate the coordinates of HSPs, setting the value of $\frac{c}{a}$ and $\gamma = 118.78°$. We obtain $T, L, U,$ and $W$ points as $\left(\frac{2\pi}{a}\right)(0,0, 0.2037)$, $\left(\frac{2\pi}{a}\right)(0.5611, 0.1620, 0.0873)$, $\left(\frac{2\pi}{a}\right)(0.5430, 0, 0.2037)$, and $\left(\frac{2\pi}{a}\right)(-0.5430, 0.3135, 0.2037)$, respectively. We use these results below in graphical representation of the single particle excitation spectrum. Here one needs to keep in mind that, depending on the conventions used for defining the reciprocal lattice vectors and the choice of the primitive cell, there may be some differences in the exact coordinates of high-symmetry points reported in different references or databases. However, the relative positions of the high-symmetry points and their symmetry properties should be consistent among different descriptions.

**B. Continuum Model**

The Hamiltonian of a general 2-level system can always be written in the basis of the three Pauli matrices($\boldsymbol{\sigma}$). We now present an extended version of the continuum model of WSM in refs. [20] in the basis of these matrices. This is given by

$$H'_\xi(\boldsymbol{k}) = \xi\hbar v_F a^{-1}\boldsymbol{k}\cdot\boldsymbol{\sigma} + (\xi\hbar a^{-1}\boldsymbol{\beta}_\xi\cdot\boldsymbol{k} - \mu)\sigma_0 + P(\boldsymbol{k})\sigma_z - \xi Q\sigma_z + \boldsymbol{M}(\eta_1, \eta_2, \theta)\cdot\boldsymbol{\sigma},$$
(1)

where $P(\boldsymbol{k}) = P_0 - P_1(k_x^2 + k_y^2) - P_1 k_z^2$ with $\boldsymbol{k} = a\left(k_x, k_y, \frac{c}{a}k_z\right)$, $(P_0, P_1)$ are the relevant material dependent parameters in units of energy, and $\mu$ is the chemical potential of the fermion number. While $\xi = \pm 1$ specifies the chirality of the Weyl points, the energy parameter $Q$ determines the shift of the Weyl nodes. The Pauli matrices $\boldsymbol{\sigma}$ are acting in the space of bands that make contact at Weyl point pairs. We take the Fermi velocity $v_F$ to be positive. We have included the magnetic moment (clubbed with the exchange field) $\boldsymbol{M}(\eta_1, \eta_2, \theta)$ in order to take care of the ferromagnetic order. The time reversal (TR) operator for a spin 1/2 particle is $\Theta = -i\sigma_y K$. The operator $K$ stands for the complex conjugation. We find that $\Theta H'_\xi(\boldsymbol{k})\Theta^{-1} \neq H'_\xi(-\boldsymbol{k})$, i.e. the time reversal symmetry(TRS) is not obeyed. We shall on focus on neighborhood of HSPs $T, L, U,$ and $W$ for graphical representations. For $T$ and $U$, in particular, $ak_y = 0$, while for L and W we have $ak_y = ak_{y0} = 0.1620$ and $0.3135$ respectively. Thus, one can then think of the Hamiltonian involving the wave vector components $(ak_{z,x}, 0/ak_{y0})$ as an effective two-dimensional Hamiltonian in momentum space for the investigation purpose. The magnetic moment, in general, may be given by $\boldsymbol{M} = M (\sin\psi \cos\varphi, \sin\psi \sin\varphi, \cos\psi)$ in spherical polar coordinates. However, we assume that the spin structure depends on the angle θ formed by the spin moments and the y-axis perpendicular to the the z-x plane of the sample with length, breadth, and width along $z$-,$x$-,and $y$-axes, respectively. In that case, one may write $\boldsymbol{M} (\eta_1, \eta_2, \theta) = M (\eta_1 \sin\theta, \eta_2 \sin\theta, \cos\theta)$, where $\eta_1^2 + \eta_2^2 = 1$. While θ = 0 corresponds to the ferromagnetic order along the axis perpendicular to the z-x plane, θ = 90° corresponds to the in-plane spin order. Apart from the former case, our focus will also be on the latter case. The reason being the FM materials with in-plane magnetization has not been studied so extensively compared to those with the out of the plane magnetization. The tilt of the Weyl cones is described by the parameters $\boldsymbol{\beta}_\xi$ — the tilt velocity. We shall mainly focus on the case $\boldsymbol{\beta}_\xi = \beta_\xi \boldsymbol{e}$, where the unit vector $\boldsymbol{e}$ is assumed to be in the x-z plane. Therefore, one may write $\boldsymbol{e} = (\cos\psi, 0, \sin\psi)$. It must be noted here the strain can modify a pre-existing tilt making it inhomogeneous through the sample. Upon dividing both the sides of Eq.(1) by $(\hbar v_F a^{-1})$, one obtains the dimensionless Hamiltonian

$$H_\xi(\boldsymbol{k}, Q) = \xi v_F a^{-1}(\boldsymbol{k} - \xi e \boldsymbol{A}).\boldsymbol{\sigma} + (\xi v_F a^{-1}\boldsymbol{\alpha}_\xi.\boldsymbol{k} - \mu)\sigma_0 + P(\boldsymbol{k})\sigma_z - \xi Q\sigma_z \qquad (2)$$

where $\hbar = 1, \boldsymbol{A} = -\frac{M(\eta_1,\eta_2,\theta)}{ev_F a^{-1}}$, and $\boldsymbol{\alpha}_\xi = \frac{\boldsymbol{\beta}_\xi}{v_F}$. The parameter $\boldsymbol{\alpha}_\xi$ is the dimensionless tilt velocity. Obviously enough, the magnetic moments in our general structure is equivalent to the chiral gauge field $\boldsymbol{A}$. Furthermore, one can write the second term $(\xi \boldsymbol{\alpha}_\xi.\boldsymbol{k}\sigma_0)$ as $\xi \alpha_\xi \boldsymbol{e}.\boldsymbol{k}\sigma_0$. We notice that, in the absence of the tilt, the Hamiltonian obeys $PH_\xi(\boldsymbol{k})P^{-1} = -H_\xi(-\boldsymbol{k})$ where the particle-hole symmetry(PHS) operator $P = \sigma_x K$. Thus, the tilt factor breaks PHS. Whereas for the type-I WSM the tilt factor $|\alpha_\xi| < 1$, for the type-II variety it is greater than one. We present the investigation of the model Hamiltonian (2) in section 3. The investigation may serve as a platform for realizing large anomalous Nernst effect.

**C. Lattice Model**

In the kagome compound $Co_3Sn_2S_2$, an important feature is that the $p$ orbitals of post-transition metal (Sn) atoms overlap and hybridize with the transition metal (Co) $3d$ orbitals near the Fermi level. In fact, the states at the Fermi level are comprised primarily of Co-3d orbitals. Therefore, an archetypal Hamiltonian for the compound $Co_3Sn_2S_2$ must involve the $p$-$d$ hybridization. Also, there have to be a term representing ferromagnetism of Co atoms, and the Kane-Mele type spin-orbit coupling ($t_{soc}$) involving the second-nearest-neighbour intralayer lattice vectors of a Kagome layer ($\boldsymbol{d_1, d_2, d_3}$); this term is zero when for the first neighbor (the lattice vectors are ($\boldsymbol{b_1, b_2, b_3}$)) is involved. The origin of this term can be traced back to the Sn atom residing on the center of each hexagon in the kagome lattice. Furthermore, there are the hopping integrals, viz. the first-neighbor hopping ($t_1$) and the second-neighbour hopping ($t_2$) in the Kagome layer. It must be mentioned that NN and NNN hopping terms, which describe the interactions between Co atoms in the Kagome lattice plane, were found to be inadequate for capturing the nodal line feature**[16,21]**. Therefore, the inter-Kagome-layer hopping ($t_z$) needs to be included. Upon introducing the term ($M\sigma_z$) describing the ferromagnetic ordering within the mean field approximation, the onsite energy of $p$ orbital and $d$ orbitals, ($\epsilon_p, \epsilon_d$), and the hopping integral ($t_{pd}$) between nearest Co and Sn1 sites, the momentum space representation of the Hamiltonian, in the basis ($d_{k,A,\tau=\uparrow,\downarrow}$ $d_{k,B,\tau}$ $d_{k,C,\tau}$ $p_{k,\tau}$)$^T$ where $d_{k,A,\tau}$ and $p_{k,\tau}$, respectively, are the destruction operators corresponding to $d$- and $p$-electrons, $\tau$ is the spin index and the (A,B,C) are sub-lattice indices, is expressed as $\hbar = \hbar_1 + \hbar_2$. Here

$$\hbar_1 = M \begin{pmatrix} \sigma^{12} & 0 \\ 0 & \frac{1}{2}(\sigma^{12} + \gamma^3 \gamma^5) \end{pmatrix} + \varepsilon \begin{pmatrix} \mathbb{I} & 0 \\ 0 & \frac{1}{2}(\mathbb{I} + \gamma^0) \end{pmatrix} + (\epsilon_p - \mu) \begin{pmatrix} 0 & 0 \\ 0 & \frac{1}{2}(\mathbb{I} - \gamma^0) \end{pmatrix}, \quad (3)$$

$$\hbar_2 = A_1' \begin{pmatrix} \gamma^5 & 0 \\ 0 & 0 \end{pmatrix} + i A_1'' \begin{pmatrix} \gamma^3 & 0 \\ 0 & 0 \end{pmatrix} + A_2' \begin{pmatrix} 0 & \frac{1}{2}(\mathbb{I} + \gamma^0) \\ \frac{1}{2}(\mathbb{I} + \gamma^0) & 0 \end{pmatrix}$$

$$-iA_2'' \begin{pmatrix} 0 & \frac{1}{2}(\sigma^{12} + \gamma^3 \gamma^5) \\ -\frac{1}{2}(\sigma^{12} + +\gamma^3 \gamma^5) & 0 \end{pmatrix} + A_3' \begin{pmatrix} 0 & \frac{1}{2}(\gamma^5 - \gamma^0\gamma^5) \\ \frac{1}{2}(\gamma^5 + \gamma^0\gamma^5) & 0 \end{pmatrix}$$

$$-iA_3'' \begin{pmatrix} 0 & \frac{1}{2}(\gamma^3 - \gamma^0 \gamma^3) \\ \frac{1}{2}(\gamma^3 + \gamma^0 \gamma^3) & 0 \end{pmatrix} + B_1 \begin{pmatrix} 0 & \frac{1}{2}(\gamma^5 + \gamma^0 \gamma^5) \\ \frac{1}{2}(\gamma^5 - \gamma^0 \gamma^5) & 0 \end{pmatrix}$$

$$+ B_2 \begin{pmatrix} 0 & \frac{1}{2}(\mathbb{I} - \gamma^0) \\ \frac{1}{2}(-\mathbb{I} + \gamma^0) & 0 \end{pmatrix} + B_3 \begin{pmatrix} 0 & 0 \\ 0 & \gamma^0 \gamma^5 \end{pmatrix} \quad (4)$$

Here $\mathbb{I}$ is the $4 \times 4$ identity matrix, $\sigma^{\nu\rho} = \left(\frac{i}{2}\right)[\gamma_\nu, \gamma_\rho]$, and the anti-commuting ($\frac{1}{2}\{\gamma^\mu, \gamma^\nu\} = g^{\mu\nu} \mathbb{I}$) matrices ($\gamma^0, \gamma^1, \gamma^2, \gamma^3, \gamma^5$) are Dirac matrices in contravariant notations. We note that the $\gamma^0$ matrix is hermitian while the ($\gamma^1, \gamma^2, \gamma^3$) matrices are anti-hermitian. Furthermore,

$$A_1' = 2t_1 \cos(\boldsymbol{b}_1 \cdot \boldsymbol{k}) + 2t_2 \cos(\boldsymbol{d}_1 \cdot \boldsymbol{k}) + 2t_z \cos(\boldsymbol{c}_1 \cdot \boldsymbol{k}), \quad A_1'' = 2t_{soc} \cos(\boldsymbol{d}_1 \cdot \boldsymbol{k}), \quad (5)$$

$$A_2' = 2t_1 \cos(\boldsymbol{b}_3 \cdot \boldsymbol{k}) + 2t_2 \cos(\boldsymbol{d}_3 \cdot \boldsymbol{k}) + 2t_z \cos(\boldsymbol{c}_3 \cdot \boldsymbol{k}), \quad A_2'' = 2t_{soc} \cos(\boldsymbol{d}_3 \cdot \boldsymbol{k}), \quad (6)$$

$$A_3' = 2t_1 \cos(\boldsymbol{b}_2 \cdot \boldsymbol{k}) + 2t_2 \cos(\boldsymbol{d}_2 \cdot \boldsymbol{k}) + 2t_z \cos(\boldsymbol{c}_2 \cdot \boldsymbol{k}), A_3'' = 2t_{soc} \cos(\boldsymbol{d}_2 \cdot \boldsymbol{k}), \quad (7)$$

$$B_1 = 2it_{pd}\sigma_0 \sin\left(\frac{1}{2}\boldsymbol{a}_1 \cdot \boldsymbol{k}\right), \quad (8)$$

$$B_2 = 2it_{pd}\sigma_0 \sin\left(\frac{1}{2}\boldsymbol{a}_2 \cdot \boldsymbol{k}\right), \quad B_3 = 2i\, t_{pd}\sigma_0 \sin\left(\frac{1}{2}\boldsymbol{a}_3 \cdot \boldsymbol{k}\right), \quad (9)$$

where $\varepsilon = (\epsilon_d - \mu)$. Thus, $\hbar$ is a $8 \times 8$ matrix. Here $\sigma_j's$ are the Pauli matrices and $\mu$ is the chemical potential of the fermion number. The model is described by a primitive lattice with three basis vectors: **a₁, a₂,** and **a₃** given above. In this model, the unit cell includes three Co atoms on the Kagome lattice and one Sn atom on the triangular lattice, as already mentioned.

## 3. Energy eigenspectra

The energy eigenvalues ($E_{\alpha,\xi}, \alpha = \pm 1$) of the matrix (1) for $\mu = 0$ are given by

$$E_{\alpha,\xi} = \alpha \left( \left( P(\mathbf{k}) + \xi \hbar v_F \frac{c}{a} k_z + M\cos\theta - \xi Q \right)^2 + \hbar^2 v_F^2 a^{-2}(a^2 k_x^2 + a^2 k_y^2) + M_1^2 \right)^{\frac{1}{2}}, \quad (10)$$

where $M_1^2 \equiv M^2 \sin^2\theta + 2\hbar v_F a^{-1} \xi M \sin\theta (\eta_1 a k_x + \eta_2 a k_y)$. Here the tilt parameter $\boldsymbol{\beta}_\xi = 0$. The 2D plots of the energy eigenvalues (of the Hamiltonian matrix in (1)) as a function of $ck_z$ are shown in Figure 2. In 2(a) $ak_x = 0.5611$, and $ak_y = 0.1620$. In (b) $ak_x = -0.5430$, and $ak_y = 0.3135$. **(a)** and **(b)** The numerical values of the other parameters used are $P_0 = 0.32, P_1 = 2.71, M=1, \mu = 0, Q = 0.8, v/a \ (v = v_F) = 0.26, \theta = 0, \eta_1 = 0$, and $\eta_2 = 1$. The angle $\theta = 0$ corresponds to the ferromagnetic order along the z-direction. The dimensionless momentum parameter $\zeta = Q/\hbar v_F a^{-1}$ determines the shift of the Weyl nodes. The bands of opposite chirality almost linearly touch each other at WNs. The vertical (horizontal) lines indicate avoided crossing momenta ( the fermi energy $E_F = 0$). The Figures 2**(c)** and 2**(d)** correspond to the in-plane spin ordering as $\theta = \pi/2$. The rest of the parameter values in these figures are the same as in Figures 2**(a)** and 2**(b).** In the two figures 2**(c)** and 2**(d),** the chemical potential lies within the band gap, i.e., the system is in the insulating state. However, since the chiral gauge field $A = -\frac{M(\eta_1, \eta_2, \theta)}{ev_F a^{-1}} \neq 0$, the time reversal symmetry (TRS) is not preserved by Eq.(2). This leads to non-zero Berry curvature (BC). If one is able to show that the system in this case has access to the integer values of the chern number ($C$), one can conclude that the system is in the anomalous quantum Hall insulating state. This exercise will be taken up in section 4.

We now consider the case of the non-zero tilt parameter ( $|\alpha_\xi| \neq 0$ ). The energy eigenvalues of (2) are given by

$$E_{\alpha,\xi} = \Upsilon(k) + \alpha \left[ \left( P(\mathbf{k}) + M\cos\theta + \xi \hbar v_F \frac{c}{a} ak_z - \xi Q \right)^2 + v_F^2 a^{-2}(a^2 k_x^2 + a^2 k_y^2) + M_1^2 \right]^{1/2},$$

(11)

where $\alpha = \pm 1$, $\Upsilon(k) \equiv \xi v_F a^{-1} |\alpha_\xi| ( ak_x \cos\psi + ck_z \sin\psi)$ and $M_1^2 \equiv M^2 \sin^2\theta + 2\hbar v_F a^{-1} \xi M \sin\theta \times (\eta_1 ak_x + \eta_2 ak_y)$. The corresponding eigenvectors are given in Appendix A. These will be required while calculating the anomalous Nernst conductivity $\alpha_{xy}$.

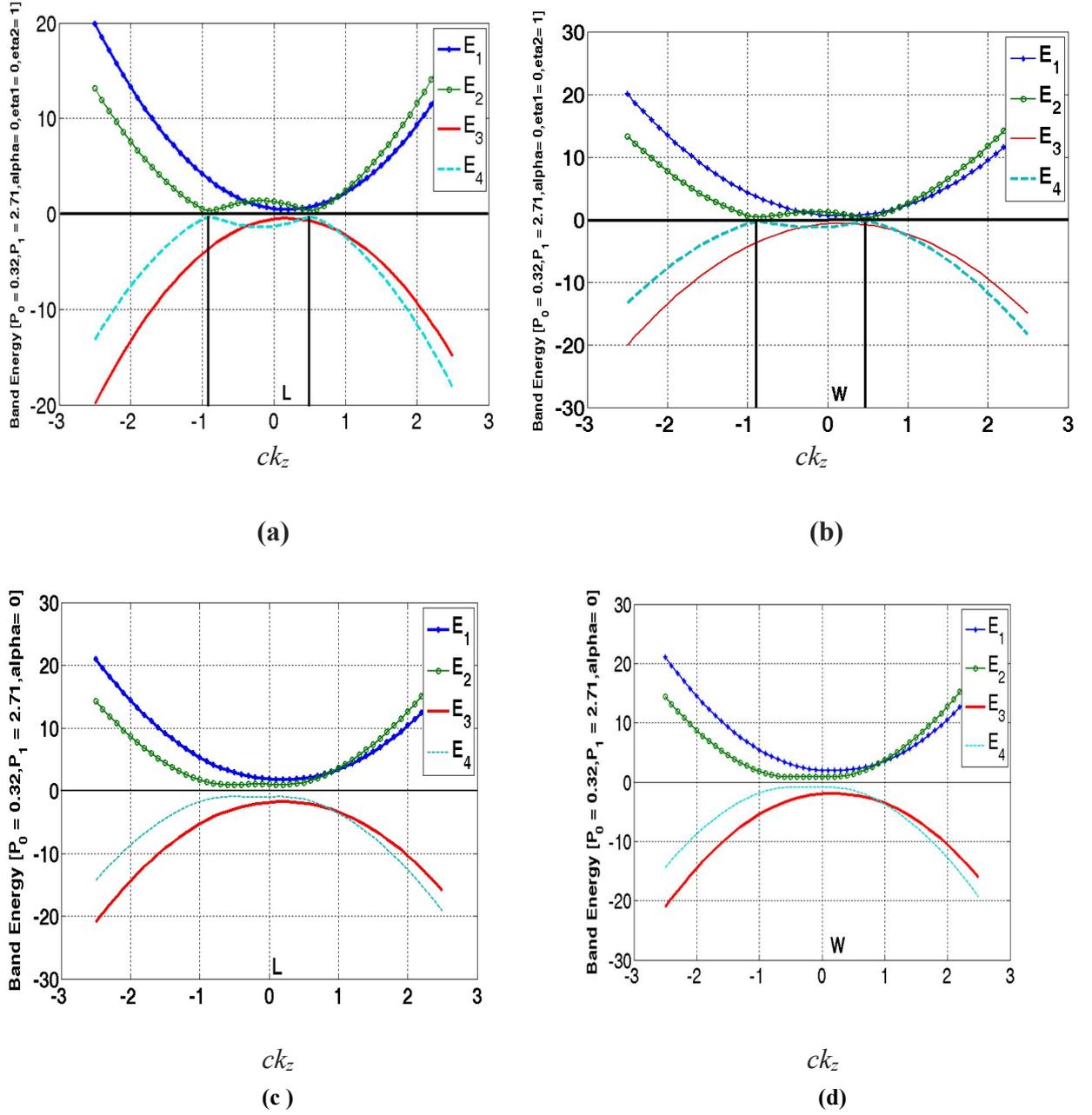

**Figure 2.** Plots of the energy eigenvalues (of the Hamiltonian matrix in (1)) as a function of $ck_z$ (the tilt parameter $\beta_\xi = 0$) around the HSPs L and W. In (**a**) and (**c**) $ak_x = 0.5611$, and $ak_y = 0.1620$. In (**b**) and (**d**) $ak_x = -0.5430$, and $ak_y = 0.3135$. (**a**) and (**b**) The numerical values of the other parameters used are $P_0 = 0.32$, $P_1 = 2.71$, $M = 1$, $\mu = 0$, $Q = 0.8$, $v/a\ (v = v_F) = 0.26$, $\theta = 0$, $\eta_1 = 0$, and $\eta_2 = 1$. The bands of opposite chirality almost linearly touch each other at WNs. The angle $\theta = 0$ corresponds to the ferromagnetic order along the z-direction. The dimensionless momentum parameter $\zeta = Q/\hbar v_F a^{-1}$ determines the shift of the Weyl nodes. The vertical (horizontal) lines indicate avoided crossing momenta ( the fermi energy $E_F = 0$). The Figures 2(**c**) and 2(**d**) correspond to the in-plane spin ordering as $\theta = \pi/2$. The rest of the parameter values in these figures are the same as in Figures 2(**a**) and 2(**b**). In 2(**c**) and 2(**d**), the chemical potential lies within the band gap.

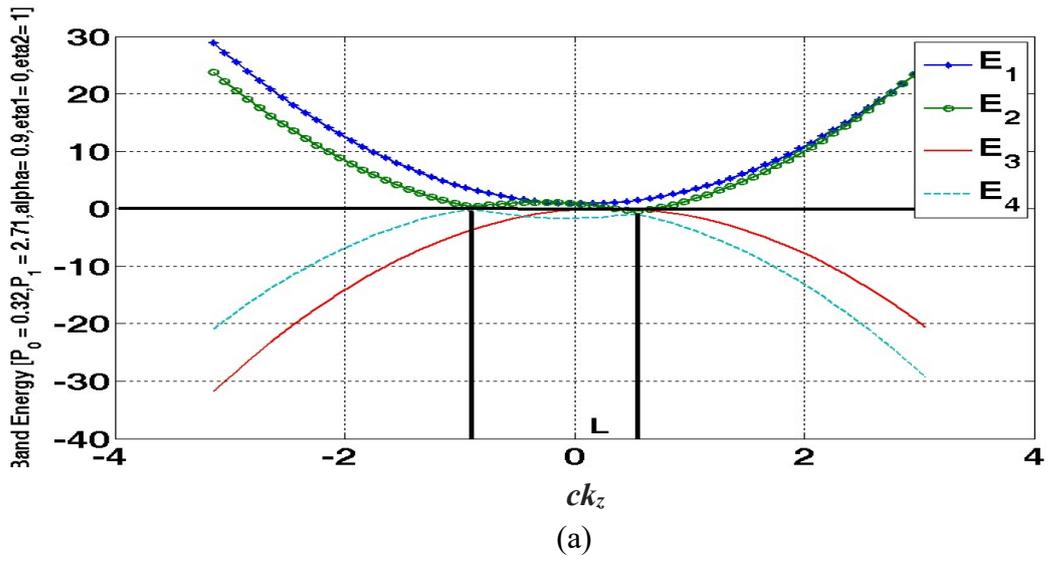

(a)

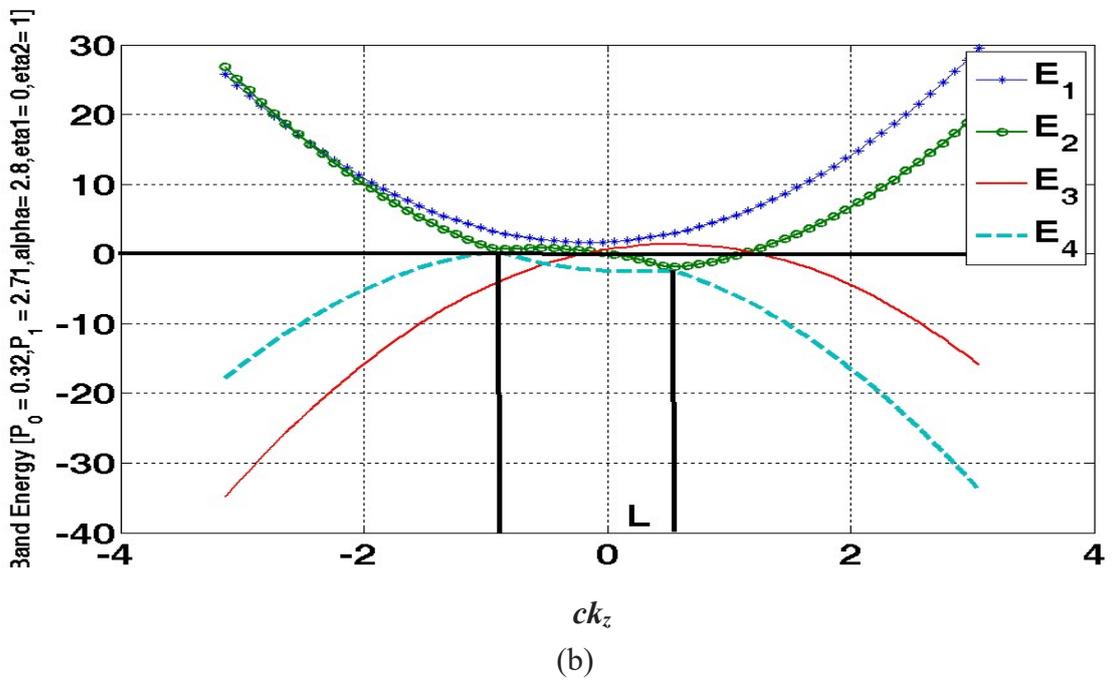

(b)

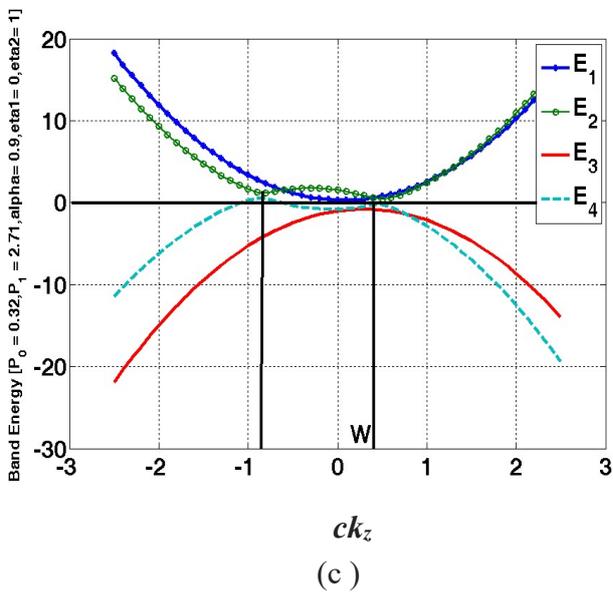

(c)

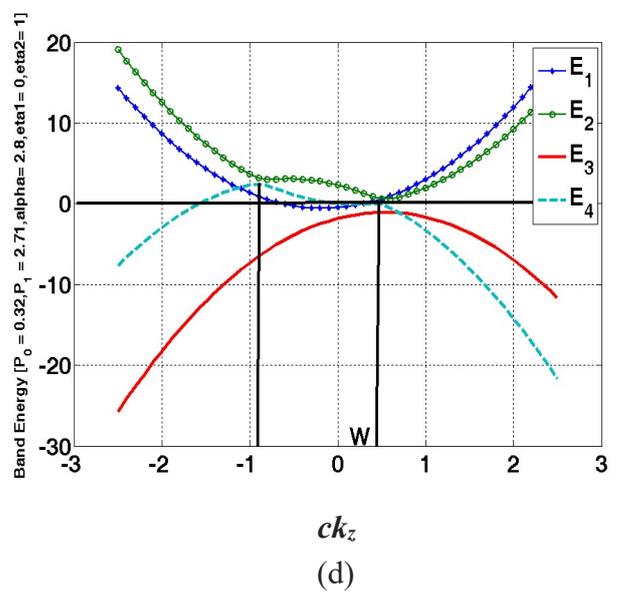

(d)

**Figure 3**. The 2D plots of energy eigenvalues in (2) as a function of $ck_z$ (around the HSPs L and W ) are shown in this figure for the tilt parameter $|\alpha_\xi|= 0.9$ ( Figures (a) and (c)) and $|\alpha_\xi|= 2.8$ ( Figures (b) and (d)). In (a) and (b) $ak_x = 0.5611$, and $ak_y = 0.1620$. In (c) and (d) $ak_x = -0.5430$, and $ak_y = 0.3135$. The numerical values of the other parameters used are $P_0 = 0.32$, $P_1 = 2.71$, $M = 1$, $\mu = 0$, $Q = 0.86$, $\theta = 0$, $v/a$ ( $v = v_F$) = 0.50, $\eta_1 = 0$, and $\eta_2 = 1$.

The 2D plots of energy eigenvalues in (2) as a function of $ck_z$ are shown in Figure 3. Here, the tilt parameter $|\alpha_\xi|= 0.8$ ( Figures (a) and (c)) and $|\alpha_\xi|= 2.8$ ( Figures (b) and (d)). The former corresponds to type-I and the latter to type-II semimetal. In (a) and (b) $ak_x = 0.5611$, and $ak_y = 0.1620$. In (c) and (d) $ak_x = -0.5430$, and $ak_y = 0.3135$. The numerical values of the other parameters used are $P_0 = 0.32$, $P_1 = 2.71$, $M = 1$, $\mu = 0$, $Q = 0.86$, $\theta = 0$, $v/a$ ( $v = v_F$) = 0.50, $\eta_1 = 0$, and $\eta_2 = 1$. We have thus shown that, for the FM order along the axis perpendicular to the plane of the system, the bands of opposite chirality almost linearly cross each other with band inversion at Weyl points above and below the Fermi level.

In order to obtain the eigenvalues of the matrix $\hbar = \hbar_1 + \hbar_2$ in Eq. (4) we need to lean on numerical analysis. Here, we choose a $k_j$-path including high symmetry point(s). For the graphical representations, the following numerical values of the various parameters were used: $t_1 = -1, t_2 = 0.6t_1$, $t_z = -1.0t_1$, $t_{pd} = 1.80t_1$, $\epsilon_d = 2.44t_1, \epsilon_p = 3.5t_1$, $\mu = 0$, and $M = 2.0t_1$. We assign the above-mentioned numerical values to the parameters appearing in the expressions ( $A'_1, A'_2, A'_3, B_1, B_2, B_3$, ……) in Eqs.(5) to (9). We use the 'Matlab' package for this purpose. Upon using the command $[V,D] = \text{eig}(\hbar_1 + \hbar_2)$ the numerical values of the eigenenergy and the corresponding eigenvectors are obtained. The command returns diagonal matrix D of eigenvalues and matrix V whose columns are the corresponding right eigenvectors, so that $ℏ *V = V*D$. For each k-point, in the chosen $k_j$-path, this process is repeated. In Figure 4, the plots of the energy eigenvalues $E_j(k)$ ($j = 1,2,........,8$) as a function of $ak_x$, obtained in this manner, are shown for $(a)(ak_y = 0, ck_z = 0,2037)$ and $(b)(ak_y = 0.1620, ck_z = 0.0873)$ without the Kane-Mele type spin-orbit coupling (SOC). The plots show the Weyl points and the band inversion. The vertical (horizontal) lines indicate avoided crossing momenta ( the fermi energy $E_F = 0$). When $t_{soc} \neq 0 (|t_{SOC}| = 1.80)$ the anticrossing feature is present with opening of spectral gaps at some points in the first Brillouin zone(BZ) around L point( see Figure 5). It must be noted that, as shown in the refs **[16,21]**, there is emergence of the nodal lines centred at the L point of BZ. The investigation of this model is presented very briefly here, as in the subsequent sections our focus will be on the continuum model given by (2). A systematic derivation of the quantities related to Berry curvature(BC) (based on this realistic model), such as the anomalous Nernst conductivity, the longitudinal magnetoconductivity, and the magneto-thermal conductivity is left for future investigation.

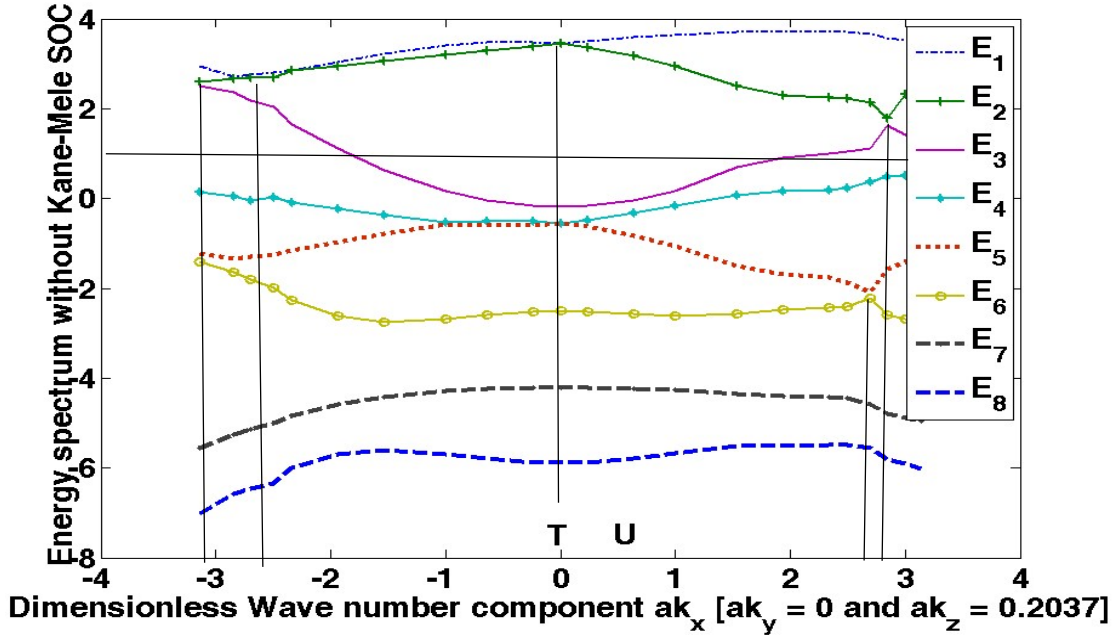

(a)

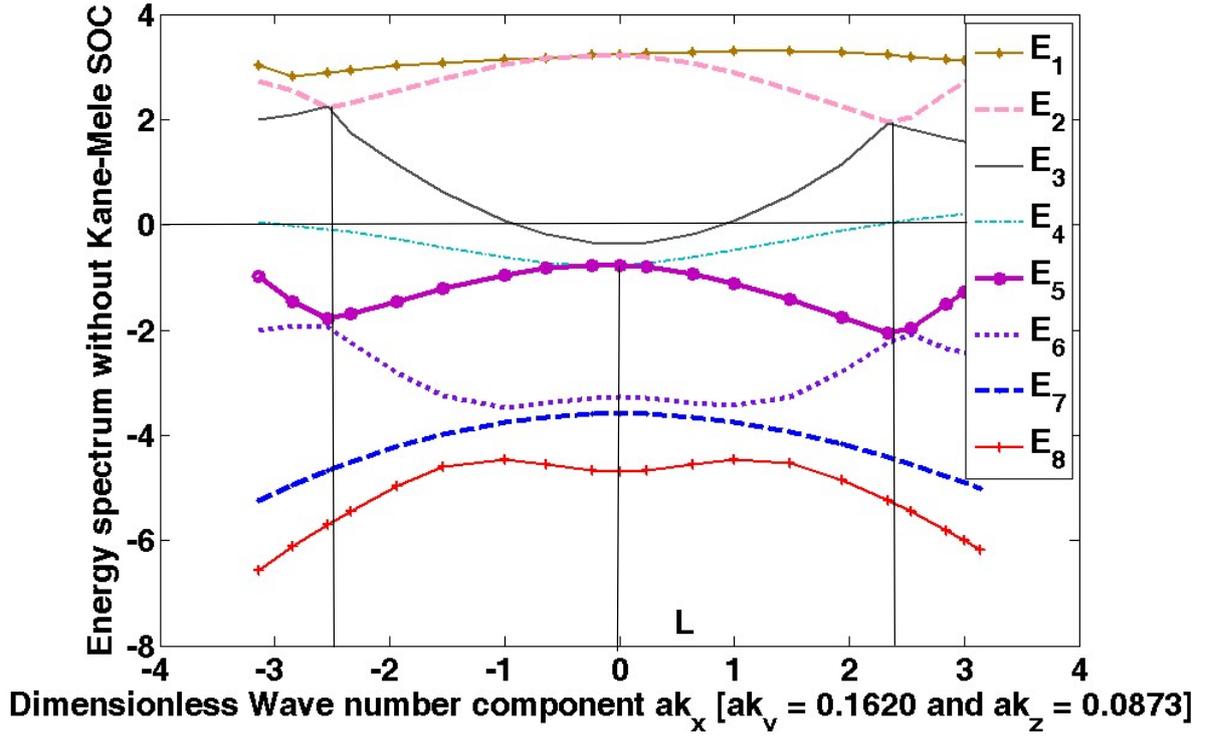

(b)

**Figure 4.** Plots of the energy eigenvalues $E_j(k)$ ($j = 1,2,........,8$) corresponding to the matrix $ℏ = ℏ_1 + ℏ_2$ in Eq.(4) as a function of $ak_x$ for $(a)(ak_y = 0, ck_z = 0,2037)$ and $(b)(ak_y = 0.1620, ck_z = 0,0873)$. We assume the following numerical values of the various parameters: $t_1 = -1, t_2 = 0.6t_1$, $t_z = -1.0t_1$, $t_{pd} = 1.80t_1$, $\epsilon_d = 2.44t_1, \epsilon_p = 3.5t_1$, $\mu = 0$, $t_{soc} = 0$, and $M = 2.0t_1$. The vertical (horizontal) lines indicate avoided crossing momenta ( the fermi energy $E_F = 0$).

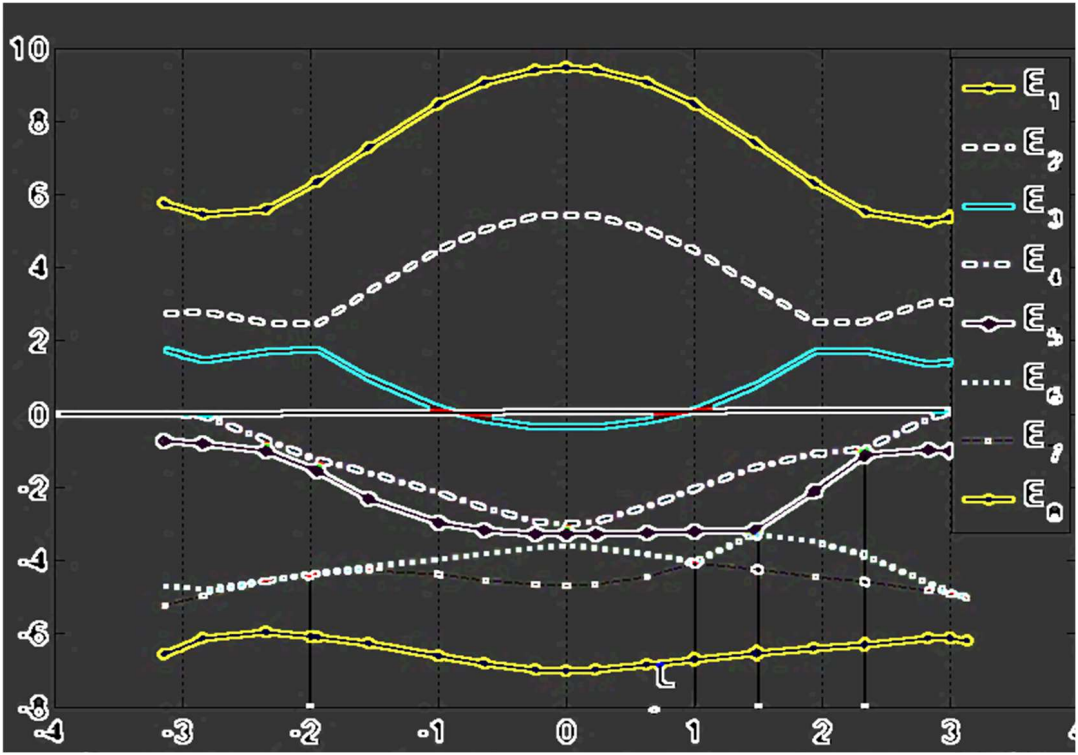

**Figure 5.** Plots of the energy eigenvalues $E_j(k)$ ($j = 1,2,\ldots\ldots,8$) corresponding to the matrix $\hbar = \hbar_1 + \hbar_2$ in Eq.(4) as a function of $ak_x$ for $(ak_y = 0.1620, ck_z = 0{,}0873)$. While the wavevector component $ak_x$ is plotted along $x$-axis, the energy eigenvalues are plotted along $y$-axis. We assume the following numerical values of the various parameters: $t_1 = -1, t_2 = 0.6t_1$, $t_z = -1.0t_1$, $t_{pd} = 1.80t_1$, $\epsilon_d = 2.44t_1$, $\epsilon_p = 3.5t_1$, $\mu = 0$, $|t_{SOC}| = 1.80$, and $M = 2.0t_1$.

## 4. Topological Properties

In this section, we present our findings on the topological properties appearing in the continuum model in (2). One of the routes to investigate the topological properties is to calculate the Berry curvature (BC), the anomalous Nernst conductivity ($\alpha_{xy}$), and the intrinsic Hall conductivity ($\sigma_{xy}$). We use the following formula **[37,38]** for calculating BC:

$\Omega_{xy}^{(n)}(k) = \sum_{\xi, m \neq n} f_{m,n,\xi,\alpha}(k, Q)$, where

$$f_{m,n,\xi,\alpha}(k, Q) = \frac{-2\, Im[\langle u_{n,\xi,\alpha}(k)|\frac{\partial H_\xi(k,Q)}{\partial k_x}|u_{m,\xi,\alpha}(k)\rangle\langle u_{m,\xi,\alpha}(k)|\frac{\partial H_\xi(k,Q)}{\partial k_y}|u_{n,\xi,\alpha}(k)\rangle]}{\{E_{n,\xi,\alpha}(k) - E_{m,\xi,\alpha}(k)\}^2}. \quad (12)$$

The function $\Omega_{xy}^{(n)}(k)$ is the $z$-component of BC for the $n^{th}$ occupied band $E_{n,\xi,\alpha}(k)$. BC is the analogue of the magnetic field in momentum-space while the Berry connection $\mathbf{A}_n(k)$ acts as a vector potential; that is, $\nabla_k \times \mathbf{A}_n(k) = \mathbf{\Omega}_n(k)$. The intrinsic Hall conductivity ($\sigma_{xy}$) is given by considering BC of all bands below the Fermi level ($E_F$): $\sigma_{xy} =$

$\frac{e^2}{\hbar} \nu$, and $\nu = \frac{1}{2\pi} \sum_{E_n \leq E_F} \int d\mathbf{k}\ \Omega_{xy}(k)$ where the integral is over the entire Brillouin zone(BZ). As shown in Figures 2(c) and 2(d) corresponding to the in-plane spin ordering ( $\theta$ = $\pi/2$), the system is an insulator with Fermi level ($E_F$) situated in the gap between the valence and the conduction band. In these cases the chern number $C = \nu$ is expected to have integer values. It may be noted that the Chern number is a property of a material and is particularly important because, being an integer, it cannot be changed under continuous deformations of the system.

The Heisenberg equation of motion is $i\hbar \frac{d\hat{x}}{dt} = [\hat{x}, \widehat{H}]$. In view of this equation, we find that the identity $\hbar \langle u_{n,\xi,\alpha}(k')|\hat{v}_j|u_{m,\xi,\alpha}(k)\rangle = \left(E_{n,\xi,\alpha}(k') - E_{m,\xi,\alpha}(k)\right)\langle u_{n,\xi,\alpha}(k')|\frac{\partial}{\partial k_j}|u_{m,\xi,\alpha}(k)\rangle$ is satisfied for a system in a periodic potential and its Bloch states as the eigenstates $|u_{n,\xi,\alpha}(k)\rangle$.

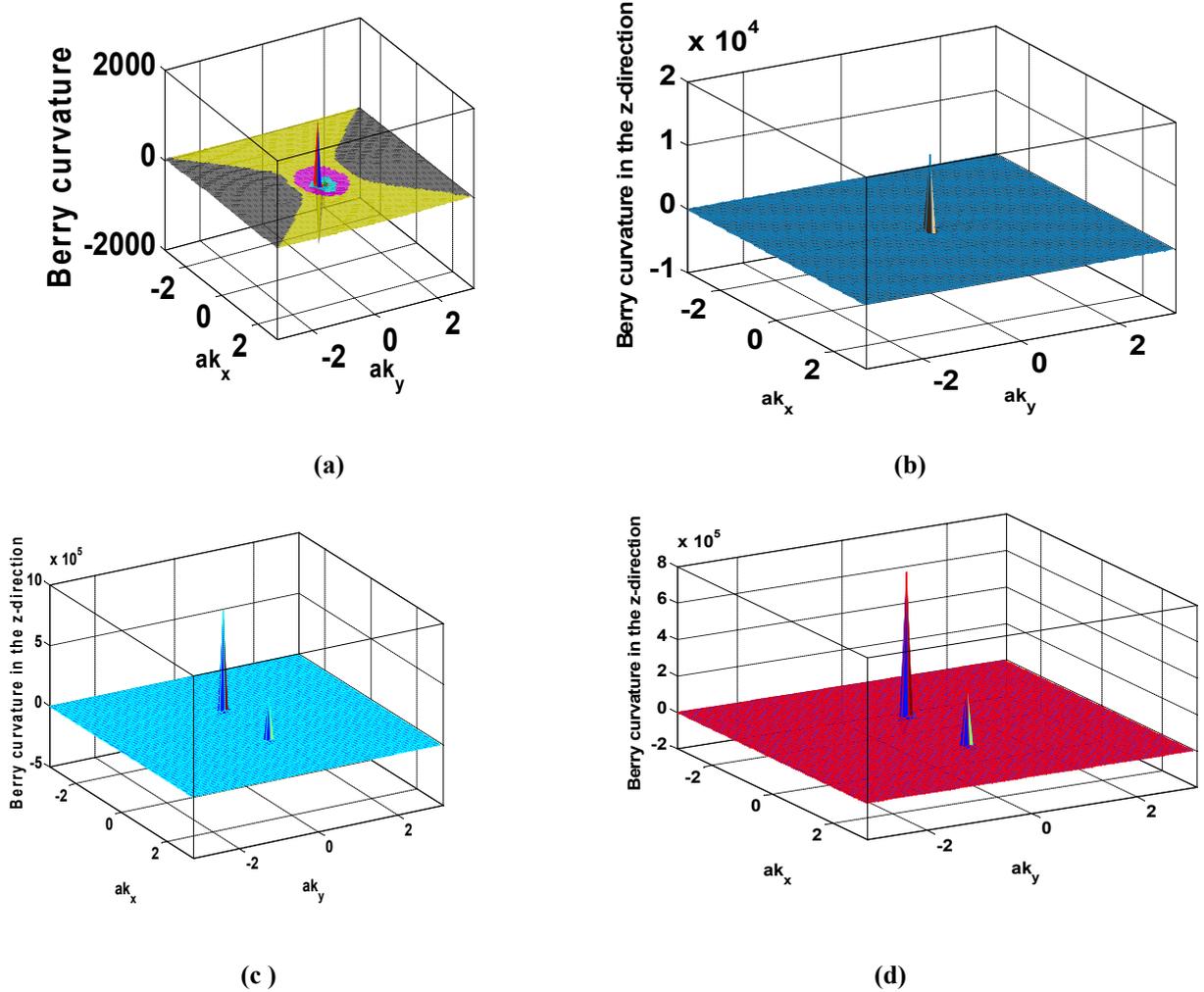

**Figure 6.** 3D plots of the Berry curvature in the z-direction as a function of $(ak_x, ak_y)$ around the HSP T(0,0,0.2037). The numerical values of the parameters used are $P_0 = 0.32$, $P_1 = 2.71$, $M=1$, $\mu = 0$, $Q = 0.86$, $v/a$ ($v = v_F$) = 0.50, $\theta = 0$, $\eta_1 = 1$, and $\eta_2 = 0$. The angle $\theta = 0$ corresponds to the ferromagnetic order along the z-direction. The tilt parameter $|\alpha_\xi| = 0.80$ ( Figures (a)) and $|\alpha_\xi| = 2.8$ ( Figures (b)). The Figures 6(c) and 6(d) correspond to the in-plane spin ordering as $\theta = \pi/2$. The rest of the parameter values in these figures

are the same as in Figures 6(a) and 6(b). (c ) Tilt parameter: 0.9, and C = 2.0997. (d) Tilt parameter: 1.14, and C = 2.0229.

Here the operator $\hbar^{-1}\frac{\partial H_\xi(k,Q)}{\partial k_j} = \hat{v}_j$ represents the velocity in the $j = (x,y)$ direction. The z-component of BC may now be written in the form $\Omega_{xy}(k) = -2\sum_n Im \left\langle \frac{\partial u_{n,\xi,\alpha}(k)}{\partial k_x} \Big| \frac{\partial u_{n,\xi,\alpha}(k)}{\partial k_y} \right\rangle$ upon using the identity above. We use this formula to calculate BC (see also Appendix A) below.

The 3D plots of the Berry curvature in the z-direction as a function of $(ak_x, ak_y)$ around the HSP T(0,0,0.2037) are shown in Figure 6. The numerical values of the parameters used are $P_0 = 0.32$, $P_1 = 2.71$, $M = 1$, $\mu = 0$, $Q = 0.86$, $v/a$ ( $v = v_F$) = 0.50, $\theta = 0$, $\eta_1 = 1$, and $\eta_2 = 0$. The tilt parameter $|\alpha_\xi| = 0.80$ ( Figures (a)) and $|\alpha_\xi| = 2.8$ ( Figures (b)). The Figures 6(c) and 6(d) correspond to the in-plane spin ordering as $\theta = \pi/2$. The rest of the parameter values in these figures are the same as in Figures 6(a) and 6(b). While in (c) tilt parameter: 0.9, and the chern number obtained is C = 2.0997, in (d) tilt parameter: 1.14, and C = 2.0229. The value of C ≈ 2 indicates that, for $\theta = \pi/2$, the anomalous Hall conductivity can be quantized leading to a topological phase with C >1.

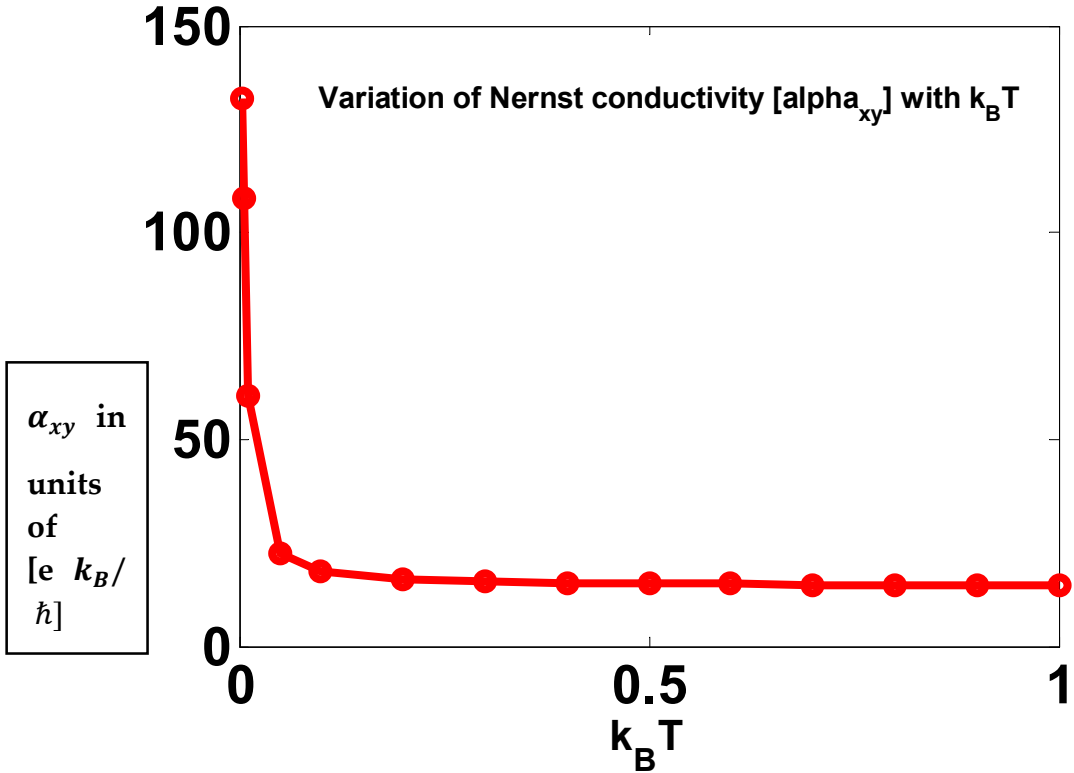

**Figure 7.** The plot of $\alpha_{xy}(T)$ as a function of $k_B T$ for $\theta = 0$. The plot shows that near $T = 0$, $\alpha_{xy}(T)$ assumes much larger value compared to those at higher temperature. The numerical values of the parameters used are the same as those in Figure 6.

In order to investigate Berry-phase effect in anomalous thermoelectric transport, we will now consider the anomalous Nernst conductivity(ANC) $\alpha_{xy}(T)$. The conductivity can be computed by integrating Berry curvature with entropy density over first BZ [39], i.e. $\alpha_{xy}(\mu, T) = k_B \frac{e}{\hbar} \sum_{n,\xi=\pm 1, \alpha=\pm 1} \int d\mathbf{k}\, \Omega_{xy}^{(n)}\, s(E_{n,\xi,\alpha}(k))$ where the entropy density $s(E_{n,\xi,\alpha}(k))$ is given by the expression

$$s(\varepsilon) = \frac{\varepsilon - \mu}{k_B T} f(\varepsilon) + \log\left(1 + \exp\left(\frac{\mu - \varepsilon}{k_B T}\right)\right), f(\varepsilon) = \frac{1}{1 + \exp\left(\frac{\varepsilon - \mu}{k_B T}\right)}. \tag{13}$$

Here, $\mu$ is the chemical potential, and $f(\varepsilon)$ is the Fermi-Dirac distribution with $k_B$ the Boltzmann constant. In the low-temperature limit, upon using the Mott relation[40] we obtain

$$\alpha_{xy}(\mu) \approx \frac{\frac{\pi^2}{3} k_B^2 T}{e} \sum_{n,\xi=\pm 1, \alpha=\pm 1} \int d\mathbf{k}\, \Omega_{xy}^{(n)} \frac{\partial f(E_{n,\xi,\alpha}(k))}{\partial E_{n,\xi,\alpha}(k)}. \tag{14}$$

The 2D graphical representation of $\alpha_{xy}(T)$ as a function of $k_B T$ is shown in Figure 7. The plot shows that near $T = 0, \alpha_{xy}(T)$ assumes much larger value compared to those at higher temperature. The result is qualitatively same as that in refs.[41,42].

## 5. Discussion and Conclusion

The model in Eq.(2) investigated in this communication to obtain a large anomalous Nernst conductivity. Also, we have shown that for the in-plane spin order (θ = π/2) in a generic FMWSM, the value of the Chern number C ≈ 2. This indicates that, for θ = π/2, the anomalous Hall conductivity can be quantized topological phase with Chern number > 1.. These are the highlights of the paper.

The model (2) can be easily extended to a low-energy, minimalistic Hamiltonian for a two-band charge-n (positive integer)Weyl point or a multi-Weyl semimetal system. All that one needs to do is to replace the off-diagonal terms in (2) by $[\,(\xi\hbar v_F a^{-1})(aK_x \mp iaK_y)^n\,]$ where $aK_x = (ak_x - \xi e A_x)$ and $aK_y = (ak_y - \xi e A_y)$. In actual lattice systems, it is possible to consider n = 1,2, and 3 [43,44]. For example, one may consider the following two-band momentum space model for a charge-2 WSM

$$\begin{pmatrix} A(k_x, k_y, k_z) & C(k_x, k_y) - iD(k_x, k_y) \\ C(k_x, k_y) + iD(k_x, k_y) & B(k_x, k_y, k_z) \end{pmatrix}$$

with broken inversion, time-reversal and mirror symmetry, where $A = 2\Gamma(P - cosk_x - cosk_y - cosk_z) + \beta\,\Gamma sink_z - \mu$, $B = -2\Gamma(P - cosk_x - cosk_y - cosk_z) + \beta\,\Gamma sink_z - \mu$, $C = -2\Gamma(cosk_x - cosk_y)$, and $D = 2\Gamma(sink_x\, sink_y)$. The numerical values of the parameters $\Gamma, P, \mu, etc.$ could be specified in conformity with ref. **[45]**. ) This model has nodes at $(k_x, k_y, k_z) = (0, 0, \cos^{-1}(P - 2))$ for $1 \leq P \leq 3$. The calculation of anomalous Nernst conductivity could be carried out in the manner as above. As regards minimum lattice model of Charge-four weyl point and its chirality-dependent properties, one may refer to ref.**[46]**.

As mentioned in section 2B above, the strain tensor $u_{ij} = \left(\frac{1}{2}\right)(\partial_i u_j + \partial_j u_i)$, where $u_j$ is the phonon vector field, can modify a pre-existing tilt making it inhomogeneous through the sample. In such a general case, the tilt velocity $\alpha_i$ in Hamiltonian (2) needs to be replaced by $\tilde{\alpha}_i = \alpha_i + u_{ij} \alpha_j$. This issues will be investigated in future. Also, it will be shown that in the case of WSM the Kerr and ellipticity angles are substantially increased in comparison with those for an ordinary, non-topological metal in the absence of magnetic field. Furthermore, the FM materials with the out-of-plane magnetization has been the subject of numerous studies due to their applications in high-density magnetic recording and spintronic devices**[ 47,48]**. The FM materials with in-plane magnetization has not been studied so extensively. This is one of the main motivations to make a preliminary study of FM materials with the in-plane magnetization ( see Figures 2(c), 2(d), 6(c) and 6(d)).
.
We have shown SFA in Figure 1(b). In order to explain this chirality preserving deformation of WNs we consider a minimalistic Hamiltonian ( a variant of (2) without the tilt parameter) of the form $H_\xi(\boldsymbol{k}) = \xi v_F a^{-1}(\boldsymbol{k} - \xi e \boldsymbol{K}(z)) \cdot \boldsymbol{\sigma} + (\xi v_F a^{-1} k_z - \mu)\sigma_z$ where $\boldsymbol{\sigma} = (\sigma_x, \sigma_y), \boldsymbol{k} = (k_x, k_y)$ and the vector potential $\boldsymbol{K}(z) = e^{-1}\left(aK_x(z), aK_y(z)\right)$. Obviously enough, one may define a pseudo-magnetic field here of the form $\boldsymbol{B}(z) = \boldsymbol{\nabla}_{aK} \times \boldsymbol{K}(z)$. Upon Taylor expanding around a minimum $z_{0,n}$, the corresponding pseudo Landau levels are given by $E_n = \xi v_F a^{-1}(\boldsymbol{k} - \xi e \boldsymbol{K}(z_{0,n})) \cdot \hat{\boldsymbol{B}}$ where $\hat{\boldsymbol{B}}$ is a unit vector in the direction of $\boldsymbol{B}(z)$. The pseudo-magnetic field may be interpreted as the one which stretches WNs into $E_n$. This is equivalent to SFA —an exotic non-closed surface state – shown in Figure 1(b)**[ 49,50]**.

In conclusion, the main motivation of the present work was to develop an understanding of WSMs from their effective low-energy theory. This is based on an interesting continuum model. We have investigated here anomalous Hall and Nernst conductivities which are related to BC. The longitudinal magnetoconductivity, the magneto-optical conductivity, and the magneto-thermal conductivity also bear the signature of BC. A systematic derivation of these quantities is left for future investigation. We finally note that WSMs possess many unusual optical properties such as they can exhibit BC related giant nonreciprocal effects. They may be used to construct ultra-compact optical isolators and circulators that are essential components in optical circuits.They generate high photogalvanic effect also which can potentially overcome the Shockley-Queisser limit of traditional solar cells.

**Appendix A**

The eigenvectors corresponding to the eigenvalues $E_{\alpha,\xi}$ in (11) are given by

$$|u_{\alpha,\xi}(k,Q)\rangle = \left(\frac{1}{\sqrt{N_1(k,\xi,Q)}}\right) \begin{pmatrix} \frac{\Upsilon_1(k)-i\Upsilon_2(k)}{E_{\alpha,\xi}(k)-\Upsilon(k)-A_0(k)} \\ 1 \end{pmatrix}, \quad (A.1)$$

$$N_1(k,\xi,Q) = 1 + \frac{\left(\Upsilon_1(k)\right)^2 + \left(\Upsilon_2(k)\right)^2}{\{E_{\alpha,\xi}(k)-\Upsilon(k)-A_0(k)\}^2}, \quad (A.2)$$

$$\Upsilon_1(k) = \xi v_F a^{-1}(ak_x) + \eta_1 M\sin\theta, \ \Upsilon_2(k) = \xi v_F a^{-1}(ak_y) + \eta_2 M\sin\theta, \quad (A.3)$$

$$A_0(k) = \left(P(\boldsymbol{k}) + \xi\hbar v_F \frac{c}{a} k_z + M\cos\theta - \xi Q\right). \quad (A.4)$$

$$\Upsilon(k) \equiv \xi v_F a^{-1}|\alpha_\xi|(\ ak_x\cos\psi + ck_z\sin\psi). \quad (A.5)$$

We now consider the case where $|\alpha_\xi| = 0$. In this special case, the corresponding eigenvectors are given by

$$|u_{\alpha,\xi}(k,Q)\rangle = \left(\frac{1}{\sqrt{N(k,\alpha,\xi,Q)}}\right) \begin{pmatrix} \frac{\xi v_F aK_-}{\alpha[A_0^2(k)+(v_F aK)^2]^{1/2}-A_0(k)} \\ 1 \end{pmatrix} \quad (A.6)$$

$$\xi v_F aK_\pm = (\xi v_F ak_\pm + \eta_1 M\sin\theta \pm i\eta_2 M\sin\theta), \quad (A.7)$$

$$N(k,\alpha,\xi,Q) = 1 + \frac{(v_F aK)^2}{\left\{\alpha[A_0^2(k)+(v_F aK)^2]^{\frac{1}{2}}-A_0(k)\right\}^2}. \quad (A.8)$$

The eigenstates above are required for the calculation of BC. Using these states and the formula $\Omega_{xy}^{(n)}(k,\xi,\alpha) = -2\,Im\,\left\langle\frac{\partial u_{n,\xi,\alpha}(k)}{\partial k_x}\bigg|\frac{\partial u_{n,\xi,\alpha}(k)}{\partial k_y}\right\rangle$ we find that

$$\Omega_{xy}^{(n)}(k) = \sum_{\xi=\pm 1,\alpha=\pm 1}\Omega_{xy}^{(n)}(k,\xi,\alpha) = -2\sum_{\xi=\pm 1,\alpha=\pm 1}(\ A_1(k)\,B_2(k) + A_2(k)\,B_1(k)), \quad (A.9)$$

$$A_1(k) = -\frac{1}{2N_1^{\frac{3}{2}}}\frac{\partial N_1(k)}{\partial k_x}\frac{\Upsilon_1(k)}{\{E_{\alpha,\xi}(k)-\Upsilon(k)-A_0(k)\}} + \frac{\frac{\partial \Upsilon_1(k)}{\partial k_x}N_1^{-\frac{1}{2}}}{\{E_{\alpha,\xi}(k)-\Upsilon(k)-A_0(k)\}} - \frac{N_1^{-\frac{1}{2}}\Upsilon_1(k)\frac{\partial E_0(k,\alpha,\xi)}{\partial k_x}}{\{E_{\alpha,\xi}(k)-\Upsilon(k)-A_0(k)\}^2},$$

(A.10)

$$B_1(k) = -\frac{1}{2N_1^{\frac{3}{2}}}\frac{\partial N_1(k)}{\partial k_x}\frac{\Upsilon_2(k)}{\{E_{\alpha,\xi}(k)-\Upsilon(k)-A_0(k)\}} + \frac{\frac{\partial \Upsilon_2(k)}{\partial k_x}N_1^{-\frac{1}{2}}}{\{E_{\alpha,\xi}(k)-\Upsilon(k)-A_0(k)\}} - \frac{N_1^{-\frac{1}{2}}\Upsilon_2(k)\frac{\partial E_0(k,\alpha,\xi)}{\partial k_x}}{\{E_{\alpha,\xi}(k)-\Upsilon(k)-A_0(k)\}^2},$$

(A.11)

$$A_2(k) = -\frac{1}{2N_1^{\frac{3}{2}}}\frac{\partial N_1(k)}{\partial k_y}\frac{\Upsilon_1(k)}{\{E_{\alpha,\xi}(k)-\Upsilon(k)-A_0(k)\}} + \frac{\frac{\partial \Upsilon_1(k)}{\partial k_y}N_1^{-\frac{1}{2}}}{\{E_{\alpha,\xi}(k)-\Upsilon(k)-A_0(k)\}} - \frac{N_1^{-\frac{1}{2}}\Upsilon_1(k)\frac{\partial E_0(k,\alpha,\xi)}{\partial k_y}}{\{E_{\alpha,\xi}(k)-\Upsilon(k)-A_0(k)\}^2},$$

(A.12)

$$B_2(k) = \frac{1}{2N_1^{\frac{3}{2}}}\frac{\partial N_1(k)}{\partial k_y}\frac{\Upsilon_2(k)}{\{E_{\alpha,\xi}(k)-\Upsilon(k)-A_0(k)\}} - \frac{\frac{\partial \Upsilon_2(k)}{\partial k_y}N_1^{-\frac{1}{2}}}{\{E_{\alpha,\xi}(k)-\Upsilon(k)-A_0(k)\}} + \frac{N_1^{-\frac{1}{2}}\Upsilon_2(k)\frac{\partial E_0(k,\alpha,\xi)}{\partial k_y}}{\{E_{\alpha,\xi}(k)-\Upsilon(k)-A_0(k)\}^2},$$

(A.13)

$$E_0(k,\alpha,\xi) = \{E_{\alpha,\xi}(k) - \Upsilon(k) - A_0(k)\}. \tag{A.14}$$

Equations $(A.9) - (A.14)$ have been used to obtain the graphical representations of BC in Figure 6.